# Modeling of a Reconfigurable OFDM IP Block Family
# For an RF System Simulator

Hannu Heusala
Jussi Liedes
*University of Oulu, Finland*
*e-mail: firstname.surname@ee.oulu.fi*

## Abstract

*The idea of design domain specific Mother Model of IP block family as a base of modeling of system integration is presented here. A common reconfigurable Mother Model for ten different standardized digital OFDM transmitters has been developed. By means of a set of parameters, the mother model can be reconfigured to any of the ten selected standards. So far the applicability of the proposed reconfiguration and analog-digital co-modeling methods have been proved by modeling the function of the digital parts of three, 802.11a, ADSL and DRM, transmitters in an RF system simulator. The model is intended to be used as signal source template in RF system simulations. The concept is not restricted to signal sources, it can be applied to any IP block development.*

*The idea of the Mother Model will be applied in other design domains to prove that in certain application areas, OFDM transceivers in this case, the design process can progress simultaneously in different design domains - mixed signal, system and RTL-architectural – without the need of high-level synthesis. Only the Mother Models of three design domains are needed to be formally proved to function as specified.*

## 1. Introduction

The ever-growing complexity of digital communication systems has brought new difficulties for analog RF design, too. As the performance of System-on-Chip is highly dependent on digital baseband processing as well, the digital parts have to be included in RF system simulations to ensure the functionality of the system as a whole. The modeling of these complex digital parts in an RF system simulator has become burdensome.

The digital parts that need to be co-modeled in the RF system simulator are the parts that interface with the analog RF environment and parts that have any significant effect on the performance of the system. In the case of a typical digital transceiver, functionality of the whole physical layer of the transmitter and the receiver need to be modeled. Models of the digital parts can be found from current IP block markets, but the abstraction level of such IP blocks is not high enough for co-modeling purposes.

The IP blocks on the market are typically described at RT-level which causes an impractical increase to the simulation times. In RF system simulations only the behavioral functionality of the block is needed, not the exact structural architecture.

As a solution to these issues we propose a behavioral level reconfigurable OFDM IP block family described in a high-level language. Such modeling makes the digital-analog co-modeling possible and reasonable as explained in this paper. Here, the term Standard Family means the group of following ten standard specifications: 802.11a, 802.11g, ADSL, DRM, VDSL DAB, DVB, 802.16a, HomePlug 1.0, ADSL++.

## 2. Principle of the analog-digital co-modeling

The proposed design flow has four abstraction levels which are called here the System Level, Architecture Level, Logic Level and the Circuit Level. These abstraction levels are covered by design domains as follows. The System Level is covered by the System Domain, Mixed-Signal Domain and the Co-Design Domain. The Architectural level is covered by the Mixed-Signal Domain and the HDL Domain. The Logic Level and the Circuit Level are covered by the Gate/Transistor Domain and the Implementation Domain. A common description language and a simulation environment specify a design domain in this paper. For example, in the System Domain, a system level description language e.g. SystemC is used to describe the structure and the function of the system. A system simulator could be used to verify the behavior of the system blocks and the communication channels between them.

The digital-analog co-modeling takes place in the Mixed-Signal Domain where the functions of the system level blocks can be modeled. The performance-critical digital blocks from the System Domain are transferred to the Mixed-Signal Domain. The transferred blocks need to be translated into a high-level language suitable for the RF simulator. This is not supposed to be a difficult task since the source and destination languages are both high-level languages. The translation can however be burdensome because it has to be done manually and current systems tend to be complicated. This is why the idea of reconfigurable IP blocks is meaningful. The idea of having reconfigurable IP blocks, Mother Models, for certain standard families would save design time. TX and RX logic blocks in the System Domain and in the Mixed-Signal Domain could be instances of the same Mother Models, reconfigured into a certain standard to be designed for the SoC. With one flexible Mother Model, the Co-Modeling of several communication standards could be encompassed. The burdensome translation of the Mother Model has to be done only once.

The behavioral level models of the digital parts in the Mixed-Signal Domain form an executable specification for the RF designer. With these executable baseband blocks the RF designer can assure the functionality of the design at RF system level. With this approach it is possible to start the design flow of the analog RF parts as soon as the necessary high-level system blocks are embedded into the RF system simulator. The most important benefit is that the operation of the digital transceiver can be verified with proper modeling of the RF parts and the transmission channel in one simulator.

RF design flow continues traditionally from the RF system simulation. The digital design flow initiates from the behavioral level models at system level, but parameters or definitions can be derived or affirmed also from the RF system simulations at the Mixed-Signal Domain.

## 3. The modeling of signal sources in an RF simulator from the developed Mother Model

As an example, we have developed a high-level model of a reconfigurable OFDM transmitter that can be embedded into APLAC® System Simulator [1, 2]. The model is an executable bundle of code that performs the digital baseband processing of the OFDM modulation in an OFDM transmitter and it can be considered as an instance of the Mother Model in the Mixed-Signal Domain. The model was written in APLAC® Language, and it works as a digital signal source for the RF designer. The model was wrapped into an APLAC® Submodel, and in the RF system simulation it appears as a signal source block that can be used in traditional RF system simulations.

To prove the model, it was reconfigured to fulfill the OFDM modulation of three different standardized OFDM transmitters: IEEE 802.11a WLAN, multi-carrier ADSL modem and Digital Radio Mondiale (DRM). The reconfiguration of the model was achieved through a set of parameters, i.e. the changeover from a standard to another is achieved simply by changing the parameters of one Mother Model. Since the digital block was modeled at behavioral level, it was fast to simulate i.e. it had only negligible influence to the total simulation time of the whole transmitter [1].

The OFDM standard family was selected for this example of reconfigurable co-modeling because there is a suitable amount of digital signal processing in OFDM modulation and there are other interests on this multi-carrier technique, too [3, 4]. In addition, the OFDM standards have such common features that demarcate clear boundaries of the standard family suitable for the concept of the reconfigurable IP block.

To extend the design domain specific models of the OFDM standard family, Mother Models in SystemC and in VHDL have been programmed and are in testing phase at the moment. The parameterized formal descriptions of the Mother Models are under development

## 4. Conclusion

The concept of co-modeling digital reconfigurable IP blocks in an RF system simulator was presented in this paper. An example of such reconfigurable Mother Model was developed and tested in APLAC® System Simulator environment [1]. The reconfigurability concept saves design time in the case of multiple designs of closely related systems. Although the design time of the reconfigurable Mother Model is longer than the design time of individual standard specific model, the individual standards can be derived more quickly from the Mother Model than if designed separately from the start. In the case of two or more different standards this approach is time saving. The reconfigurable Mother Model should be considered as a template for certain selected systems.

The concept of reconfigurability is most applicable for systems related closely to each other. This means that the systems covered by the Mother Model need to have enough features in common. OFDM modulation was proven as such.